\newcommand{\PHCX} {(C$_4$H$_{12}$N$_2$)Cu$_2$(Cl$_{1-x}$Br$_{x}$)$_6$}
\newcommand{\PHCC} {(C$_4$H$_{12}$N$_2$)Cu$_2$Cl$_6$}

\newcommand{\musr} {$\mu^+$SR}

\documentclass[prb,twocolumn,floatfix,preprintnumbers,amsmath,amssymb,superscriptaddress]{revtex4}

\usepackage{graphicx}
\usepackage{dcolumn}
\usepackage{bm}
\usepackage{color}
\usepackage{xspace}

\begin{document}

\title{Effect of disorder on a pressure-induced $z=1$ magnetic quantum phase transition}

\author{A. Mannig}
 \email{manniga@ethz.ch}
\affiliation{Neutron Scattering and Magnetism, Laboratory for Solid State
Physics, ETH Z\"urich, Z\"urich, Switzerland.}
\affiliation{Laboratory for Muon Spin Spectroscopy, Paul Scherrer Institut, CH-5232 Villigen PSI, Switzerland.}

\author{J. S. M\"oller}
 \email{jmoeller@phys.ethz.ch}
\affiliation{Neutron Scattering and Magnetism, Laboratory for Solid State
Physics, ETH Z\"urich, Z\"urich, Switzerland.}

\author{M. Thede}
\affiliation{Neutron Scattering and Magnetism, Laboratory for Solid State Physics, ETH Z\"urich, Z\"urich, Switzerland.}
\affiliation{Laboratory for Muon Spin Spectroscopy, Paul Scherrer Institut, CH-5232 Villigen PSI, Switzerland.}

\author{D.~H\"uvonen}
\affiliation{National Institute of Chemical Physics and Biophysics, 12618 Tallinn, Estonia.}
\affiliation{Neutron Scattering and Magnetism, Laboratory for Solid State Physics, ETH Z\"urich, Z\"urich, Switzerland.}

\author{T. Lancaster}
\affiliation{University of Durham, Centre for Materials Physics, South Road, Durham, DH1 3LE, United Kingdom.}

\author{F. Xiao}
\affiliation{University of Durham, Centre for Materials Physics, South Road, Durham, DH1 3LE, United Kingdom.}

\author{R.~C.~Williams}
\affiliation{University of Durham, Centre for Materials Physics, South Road, Durham, DH1 3LE, United Kingdom.}

\author{Z.~Guguchia}
\affiliation{Laboratory for Muon Spin Spectroscopy, Paul Scherrer Institut, CH-5232 Villigen PSI, Switzerland.}

\author{R. Khasanov}
\affiliation{Laboratory for Muon Spin Spectroscopy, Paul Scherrer Institut, CH-5232 Villigen PSI, Switzerland.}

\author{E. Morenzoni}
\affiliation{Laboratory for Muon Spin Spectroscopy, Paul Scherrer Institut, CH-5232 Villigen PSI, Switzerland.}

\author{A. Zheludev}
 \homepage{http://www.neutron.ethz.ch/}
\affiliation{Neutron Scattering and Magnetism, Laboratory for Solid State Physics, ETH Z\"urich, Z\"urich, Switzerland.}

\date{\today}

\begin{abstract}

Pressure-induced ordering close to a $z=1$ quantum critical point is studied in the presence of bond disorder in the quantum spin system \PHCX\ (PHCX) by means of muon-spin rotation and relaxation. As for the pure system \PHCC, pressure allows PHCX with small levels of disorder ($x\leq 7.5\%$) to be driven through a quantum critical point separating a low-pressure quantum paramagnetic phase from magnetic order at high pressures. However, the pressure-induced ordered state is highly inhomogeneous for disorder concentrations $x>1\%$. This behavior might be related to the formation of a quantum Griffiths phase above a critical disorder concentration $7.5\%<x_{\rm c}<15\%$. Br--substitution increases the critical pressure and suppresses critical temperatures and ordered moment sizes.

\end{abstract}

 \maketitle

\section{Introduction}

Quantum magnets remain the prototypes of choice for the study of quantum-critical phenomena and quantum phase transitions (QPTs), both from a theoretical and experimental point of view~\cite{sachdev1999}. Those built from organic molecules can be tailored in composition, structure, and dimensionality and are especially susceptible to perturbations like pressure. Traditionally the main focus has been on magnetic field-induced quantum phase transitions. An important example is the study of the BEC of
magnons universality class. In gapped quantum paramagnets an applied magnetic field reduces the gap via the Zeeman effect until it vanishes at a quantum critical point (QCP) with dynamical critical
exponent~\cite{giamarchi2008} $z=2$. More recent research showed that qualitatively different soft mode transitions can be induced in gapped spin systems through a continuous change of exchange constants by the application of external hydrostatic pressure. In these transitions the spectrum is expected to be linear at the quantum critical point, and hence $z=1$. To date the only known experimental realizations of such a pressure-induced QCP are $\rm TlCuCl_3$~\cite{Tanaka2003,Oosawa2004,Rueegg2004} and \PHCC\ (PHCC)~\cite{Hong2010,Thede2014PRL,Perren2015}. Recently, a pressure-induced quantum phase transition from a gapped singlet ground state to plaquette state and ultimately an AF ordered state has also been reported~\cite{strontium_arxiv, strontium_nature} for the Shastry-Sutherland compound $\rm SrCu_{2}(BO_{3})_{2}$.

Inelastic neutron scattering experiments found the gap in PHCC to be reduced by the application of hydrostatic pressure\cite{Hong2010}. Subsequent muon-spin relaxation (\musr) experiments determined the presence of two distinct high-pressure magnetically-ordered phases above a critical pressure $p_{\rm c}\sim4.3$~kbar~\cite{Thede2014PRL}. The presence of a linearly-dispersing gapless Goldstone mode at a pressure of 9(1)~kbar was then confirmed by inelastic-neutron scattering experiments~\cite{Perren2015} and the application of pressure was shown to induce large changes in a single exchange pathway in the material leading to the closing of the gap at the QCP. In this paper we use \musr\ to study the phase diagram of bond-disordered \PHCX\ (PHCX) around this pressure-induced $z=1$ QCP.

While the problem of thermodynamic phase transitions in the presence of disorder is an old one (for an accessible review see e.g.\ Ref.~\onlinecite{Fisher1988}), that of \emph{quantum} phase transitions is very much a current area of research. Most interest in this area has focussed on $z=2$ quantum-phase transitions in the presence of bond-disorder (for a review see Ref.~\onlinecite{Zheludev2013}). Here disorder is created by randomly modifying the magnetic exchange pathways leaving the magnetic sites otherwise unperturbed. It has been applied in materials such as $\rm TlCu(Cl_{1-x}Br_{x})_3$~\cite{Oosawa2002_TlCuCl3X}, NiCl$_{2-2x}$Br$_{2x}\cdot$4SC(NH$_2$)$_2$~\cite{Yu2012_DTNX,Wulf2013_DTNX}, Sul-$\rm Cu_2(Cl_{1-x}Br_{x})_4$~\cite{Wulf2011_SulX}, and IPA-$\rm Cu(Cl_{1-x}Br_{x})_3$~\cite{Hong2010_IPAX,Nafradi2013_IPAX}, where in the pure cases excellent experimental realizations of various field-induced QPTs have been found. To date, the pressure-induced $z=1$ QCP in the presence of disorder has not been studied experimentally in gapped quantum magnets.

Disorder-free ($x=0$) PHCC is exceptionally well-characterized by a variety of techniques including x-ray and bulk measurements, elastic and inelastic neutron scattering~\cite{Stone2001,Stone2006,Stone2006Nature,Stone2007}, and electron spin resonance~\cite{Glazkov2012}. PHCC crystallizes in the triclinic space group $P\bar{1}$ with lattice parameters $a=7.984(4)\,{\rm A}$, $b=7.054(4)\,{\rm A}$, $c=6.104(3)\,{\rm A}$, $\alpha=111.23(8)^\circ$, $\beta=99.95(9)^\circ$, $\gamma=81.26(7)^\circ$. The spin--$1/2$ Cu$^{2+}$ ions are connected by a complex layered spin network, with some  degree of frustration. The ground state is a non-magnetic spin singlet with only short-range correlations. The lowest energy excitations are a $S=1$ triplet, with a gap $\Delta=1.0\,{\rm meV}$ and a bandwidth of $1.7\,{\rm meV}$.  Magnetic ordering can be induced in PHCC by application of a magnetic field~\cite{Stone2006}.

In PHCX an increasing bromine content $x$ is found to result in a linear increase of the lattice constants, which can be thought of as ``negative chemical pressure'' (e.g.,\ $c$ increases by $0.45\%$ for disorder concentration $x=12\%$)~\cite{Dan2013_xray}. Neutron diffraction and inelastic neutron scattering show that Br substitution affects the magnetic bonds instead of creating additional structural or magnetic defects~\cite{Dan2012}. The energy gap increases with Br-substitution~\cite{Dan2012_gap} to $1.5\,{\rm meV}$ for $x=7.5\%$. Thus, chemical modification increases the gap and pushes these systems away from the $z=1$ QCP and further into the quantum paramagnetic phase. No in-gap states were identified in inelastic neutron scattering, although a recent ESR absorption study suggests the formation of a local $S=1$ defect at high nominal concentrations ($x\geq5\%$) in bond-disordered PHCX~\cite{Glazkov2014,Glazkov2015}. Field-induced magnetic ordering persists up to at least $x=12.5\%$, however, the transitions become broader for higher disorder concentrations. Further findings include a reduction of magnon bandwidth and decrease of magnon lifetimes which could be attributed to the scattering of magnons by impurities. 
In this paper we study the effect of disorder on this pressure-induced $z=1$ QPT. In particular, we as the following questions: (i) Does the pressure-induced quantum phase transition exist in bond-disordered PHCX? (ii) If so, what is the nature of the pressure-induced magnetic phase?

\section{Methods}

Polycrystalline samples of $\rm( C_4H_{12}N_2)(Cu_2Cl_{6(1-x)}Br_{6x})$ with varying nominal bromine content $x=1\%$, $4\%$, $7.5\%$, and 15\% of  typical mass $800\,{\rm mg}$ were grown using the same protocol as described in reference~\onlinecite{Yankova2012}. The growth involves dissolving stoichiometric amounts of piperazine and copper(II) chloride in hydrochloric/hydrobromic acid. The `nominal bromine content' $x$ is the HBr/HCl solvent ratio in the starting solutions. The actual bromine content in the resulting crystals differs from this but fortunately the site-specific substitution of bromine is well understood~\cite{Dan2013_xray}. By averaging over the three inequivalent chlorine/bromine sites the actual  Br-concentration can be related to the nominal Br-concentration through $x_{\rm actual}=0.63(3)\ x_{\rm nominal}$. For ease of comparison with previous studies the disorder concentration $x$ in this paper refers to the nominal Br-concentration $x_{\rm nominal}$. The quality of the samples was verified by x-ray diffraction using a Bruker AXS single crystal diffractometer equipped with a cooled APEX-II detector.

Pelletized PHCX samples were loaded into MP35N and CuBe piston-cylinder clamped pressure cells, specifically designed for $\mu$SR experiments. Daphne Oil 7373 was used as pressure-transmitting medium. The pressure was applied in a hydraulic press and determined by means of ac\hbox{-}susceptibility by the pressure-dependent shift of the superconducting transition of an indium probe. The pressure cells were mounted inside a $^3$He Oxford cryostat. Measurements were carried out on the GPD instrument at the Swiss muon source at Paul Scherrer Institute~\cite{khasanov2016high}. \musr\ data were collected in zero field (ZF) and weak transverse fields (wTF) of $3\,{\rm mT}$ over a range of temperatures. ZF data is particularly sensitive to spontaneous magnetic order and provides information on the type of magnetic ground state. wTF data is useful to locate transition temperatures rapidly in order to map out phase diagrams close to zero applied field. The data were analyzed with MUSRFIT~\cite{musrfit2012}.


\section{Results and Analysis}

\subsection{Zero-field data}
\begin{figure}
\unitlength1cm
\includegraphics[width=\columnwidth]{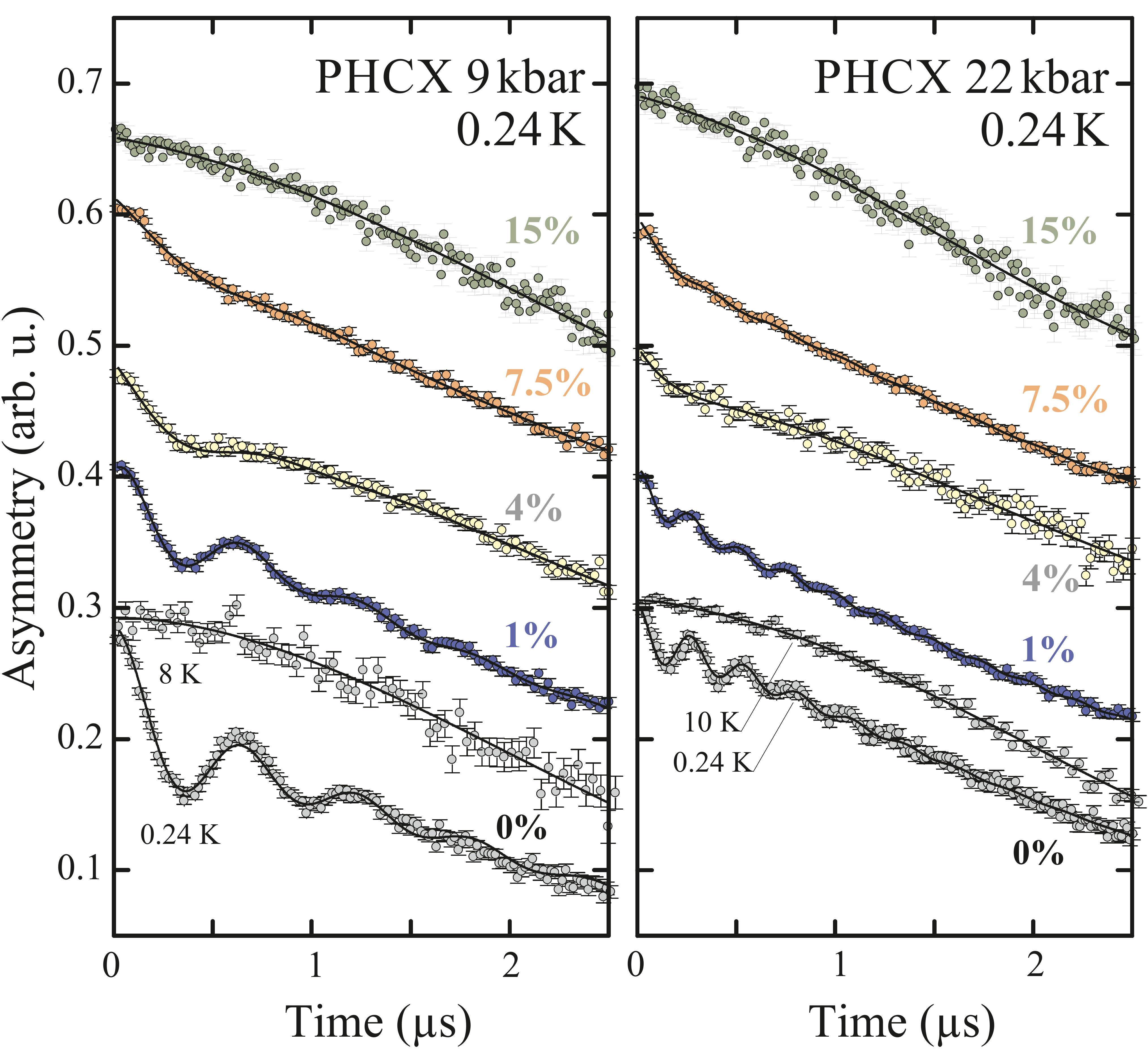}
\caption{\label{fig:PHCX_diffBrConcentr}
ZF--\musr\ spectra for PHCX for Br--concentrations of $x= 1\%$, $4\%$, $7.5\%$ and $15\%$ at intermediate and at high pressures of $9\,{\rm kbar}$ (left) and $22\,{\rm kbar}$ (right). Data were recorded generally at $0.24\,{\rm K}$, except for the $4\%$ data which were taken at $1.4\,{\rm K}$ and $2.1\,{\rm K}$ (below their ordering temperatures) for $9\,{\rm kbar}$ and $22\,{\rm kbar}$, respectively.
For comparison, pure PHCC spectra at $0.24\,{\rm K}$ and at temperatures above the ordering temperature are included. Lines are fits to the data as described in the text. The relative $y$--axis offset is $0.1$.}
\end{figure}

\begin{figure}
\unitlength1cm
\includegraphics[width=\columnwidth]{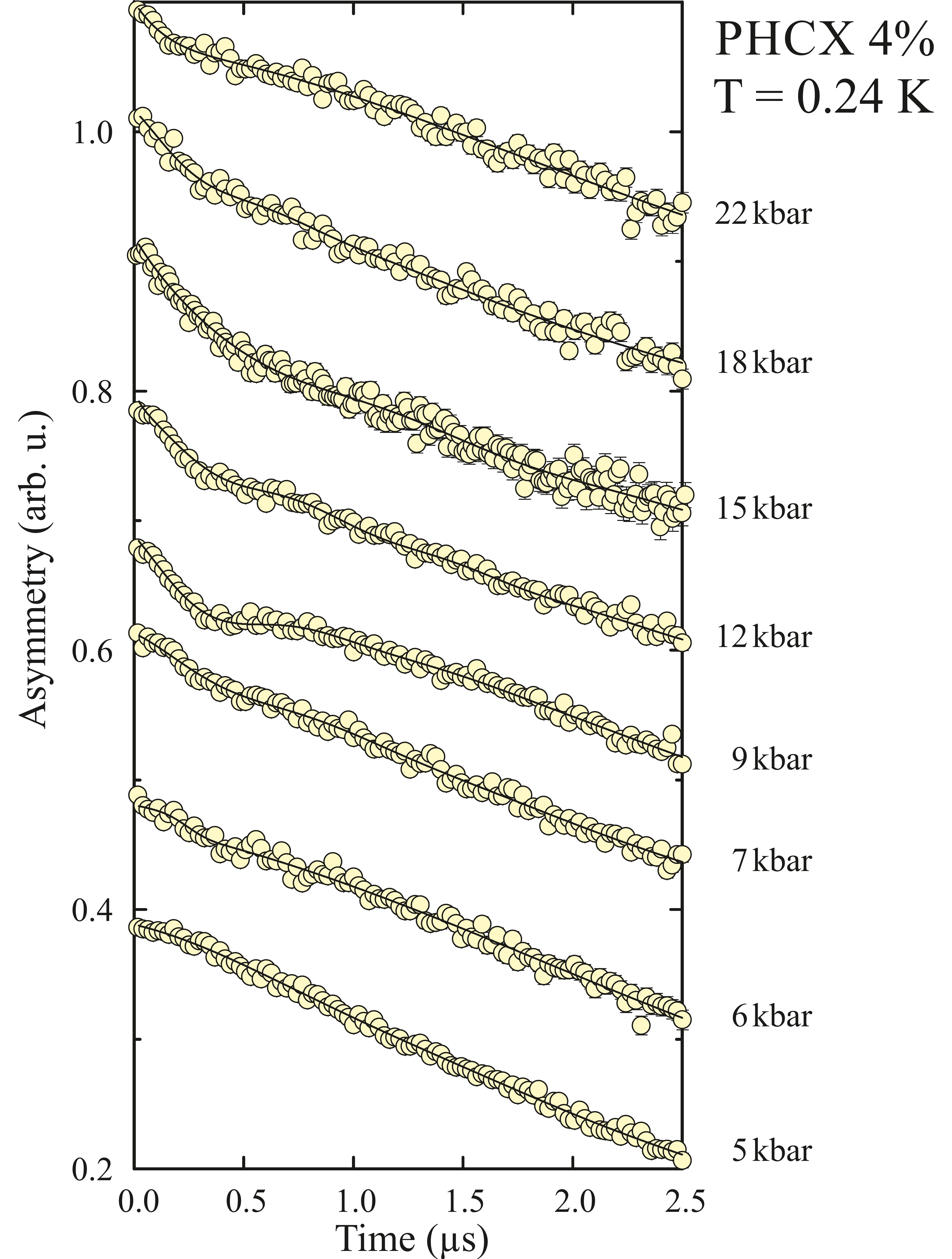}
\caption{\label{fig:PHCX_diffHydrostPress}
ZF--\musr\ spectra for $4\%$ Br--disordered PHCX at selected pressures. The data were generally recorded at $0.24\,{\rm K}$ except the data for $9\,{\rm kbar}$ and $22\,{\rm kbar}$ which were recorded at $1.4\,{\rm K}$ and $2.1\,{\rm K}$, respectively (below the ordering temperature).
Lines are fits to the data as described in the text. The relative $y$--axis offset is $0.1$.}
\end{figure}

Figure~\ref{fig:PHCX_diffBrConcentr} shows selected \musr\ spectra measured in zero applied field (ZF) for pure PHCC and Br-disordered PHCX at two different pressures~\footnote{``9\,kbar'' refers to $10.3(6)\,{\rm kbar}$, $8.6(6)\,{\rm kbar}$, $9.0(6)\,{\rm kbar}$, $9.6(6)\,{\rm kbar}$, $9.3(6)\,{\rm kbar}$ for disorder concentrations of 0\%, 1\%, 4\%, 7.5\% and 15\%, respectively, and ``22 kbar'' refers to $23.6(6)\,{\rm kbar}$, $23.0(6)\,{\rm kbar}$, $23.0(6)\,{\rm kbar}$, $22.0(6)\,{\rm kbar}$, $22.0(6)\,{\rm kbar}$} at $9\,{\rm kbar}$, and $22\,{\rm kbar}$. For comparison, the ZF spectra of pure PHCC are added.
Inevitably in a pressure experiment, there is a significant contribution to the detected signal due to muons stopping in the pressure cell walls (or cryostat tail), but fortunately the functional form of this background is well known (see below). We note that the contribution from the pressure cell is greater for higher pressures where less signal originates from the sample. Below we first qualitatively discuss the signal due to the sample before moving on to a quantitative analysis. At high temperatures, the signal due to the sample is a temperature-independent Gaussian-shaped relaxation indicative of the relaxation due to static nuclear moments only. As the temperature is lowered, we observe spontaneous oscillations in the \musr\ signal at low temperatures in pure PHCC at either of these pressures. This is evidence for a static internal magnetic field at at least one muon site due to long-range magnetic order. The ZF spectra for 1\%-disordered PHCX are very similar, although the oscillations are damped. In 4\%-disordered PHCX the oscillations are damped even more. For 7.5\%-disordered PHCX, the precession is reduced to a rapid initial depolarization. Thus, already small disorder concentrations result in a heavy damping  of muon spin precession in PHCX. This is a sign of increasingly inhomogeneous field distributions at the muon sites with increasing bromine concentrations $x$. No temperature-dependent changes in the muon signal from 15\%-disordered PHCX appear down to 0.24~K at pressures up to 22\,kbar and instead the behavior resembles that of all samples in their paramagnetic phase at high temperatures.

ZF data for 4\%--disordered PHCX are shown at a number of different pressures in Fig.~\ref{fig:PHCX_diffHydrostPress}. For $p<7$\,kbar the data show no oscillations or temperature-dependent changes that are indicative of magnetic order down to 0.24~K. However, for $p\geq 7\,\mathrm{kbar}$ damped oscillations appear at low temperatures. We hence conclude that there is a critical pressure $p_{\rm c}$ above which magnetic order occurs in the range $6~{\rm kbar}<p_{\rm c}<7$~kbar for 4\%-disordered PHCX. Above $7\,{\rm kbar}$ ZF spectra of 4\%-disordered PHCX show oscillations at low temperature with a comparable level of damping. However, at $15\,\mathrm{kbar}$ only a rapid exponential depolarization at short times can be observed.

We now turn to a detailed analysis of the ZF data. Our parameterization is similar to the one used in reference~\onlinecite{Thede2014PRL} to allow for a more direct comparison of PHCX with pure PHCC. Although other choices are possible our conclusions are unaffected. The total asymmetry $A(t)$ of all ZF spectra was described as the sum of the three components $A_{\text{s}}(t)$, $A_{\text{c}}(t)$ and $A_{\text{bg}}(t)$:

\begin{eqnarray}
A(t)=A_{\rm s}(t)+A_{\text{c}}(t)+A_{\text{bg}}(t).
\end{eqnarray} 
These components account for muons that stop in the sample,  the pressure cell, and other parts of the sample environment (pressure medium, cryostat), respectively. The respective ratios of $A_{\text{s}}(t=0)$, $A_{\text{c}}(t=0)$ and $A_{\text{bg}}(t=0)$ are independent of temperature and therefore were fitted globally for each pressure and disorder concentration. The background contribution $A_{\text{bg}}(t)$ was empirically modelled by a slowly-relaxing exponential.

\begin{eqnarray}
A_{\text{bg}}(t)=A_{\text{bg}}(t=0){\rm e}^{-\lambda_{\text{bg}}t},
\end{eqnarray}
where $\lambda_{\rm bg}\ll1$~MHz is temperature-independent and was fitted globally for a given disorder concentration and pressure.
Some 50\% to 70\% of all muons stop in the thick walls of the pressure cells but fortunately the functional form of $A_{\text{c}}(t)$ is well known to be:
\begin{eqnarray}
A_{\text{c}}(t)= A_{\text{c}}(t=0)G(t) {\rm e}^{-\lambda_{\text{c}}t},
\end{eqnarray}
where $G(t)$ is a Gaussian Kubo-Toyabe function
\begin{eqnarray}
G(t)=\tfrac{1}{3}+\tfrac{2}{3}\left[1-(\sigma t)^2\right] {\rm e}^{-\frac{1}{2}(\sigma t)^2}
\end{eqnarray}
with known temperature-independent depolarization $\sigma = 0.345\,\textrm{MHz}$, which is due to relaxation by nuclear moments.
The exponential term ${\exp}(-\lambda_{\text{c}}t)$ accounts empirically for residual dynamics in the cell. In the cells made from MP35N alloy $\lambda_{\rm c}$ is temperature-dependent below 1\,K.
Finally, the signal from the sample $A_{\text{s}}(t)$ was well-described by:

\begin{widetext}
\begin{eqnarray}
A_{\text{s}}(t)= A_{\text{s}}(t=0)\left\{y {\rm e}^{-(\lambda_{\text{para}} t)^\beta} + (1-y)\left[\tfrac{2}{3} J_0(\gamma_\mu B_{\rm m} t) {\rm e}^{-\lambda_1t} +\tfrac{1}{3}{\rm e}^{-\lambda_2t}\right]\right\}.
\end{eqnarray}
\end{widetext}

The first term between the curly brackets on the right-hand side represents the non-magnetically ordered fraction $y$. Its empirically-given parameters $\lambda_{\text{para}}$ and $\beta$ model the relaxation due to nuclear spins and electronic spin dynamics. The second term reflects the magnetic component and is modelled by a zeroth-order Bessel function of the first kind and an additional exponential relaxation. This function approximately describes the response expected from a broad field distribution with maximum field $B_{\rm m}$ at the muon site~\cite{Le1993}. The muon gyromagnetic ratio is $\gamma_{\mu} =2\pi\times135.5\,{\rm MHz/T}$. The term $\exp(-\lambda_1t)$ describes the relaxation due to additional weak dynamics. The additional exponential term with relaxation rate $\lambda_2$ models the longitudinal muon spin relaxation rate where the local field is parallel to the muon spin (the relative amplitude of  $1/3$ represents a powder average). The choice of Bessel function is desirable as it provides a good fit over the whole range of disorder concentrations and pressures. A Bessel function describes the muon depolarization in incommensurately ordered spin-density wave states~\cite{Le1993}. However, it can also provide a good empirical fit in a scenario where a range of muon sites gives rise to a broad distribution of probed fields; a situation that is likely to occur in many molecular magnets~\cite{Steele2011} as is also found by density-functional theory calculations of muon sites~\cite{Johannes_thesis,Blundell2013,Xiao2015}. Here we treat our choice of fitting function entirely empirically since our data do not allow us to distinguish between these two physically distinct scenarios. We note that a Gaussian Kubo-Toyabe function combined with an exponential relaxation provides a similarly good fit for the ZF spectra for disorder concentrations $x>1\%$.

\begin{figure}
\unitlength1cm
\includegraphics[width=\columnwidth]{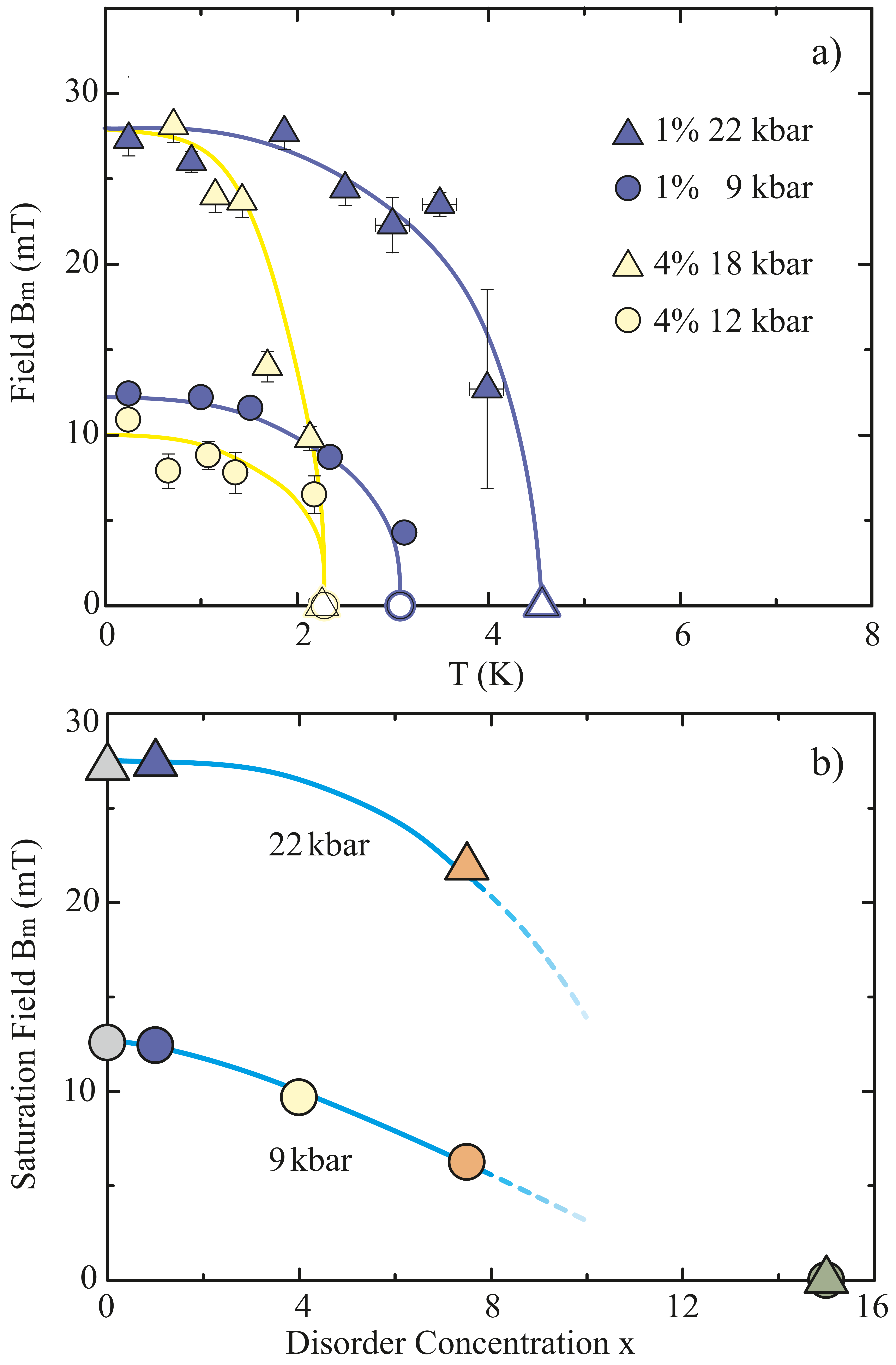}
\caption{\label{fig:EstimationSaturationField}
{ a) Thermal melting of magnetic order: temperature dependence of the fitted maximum field $B_{\rm m}$ at the muon sites for disorder concentrations of 1\% and 4\% at a selected intermediate and high pressure. Solid lines are a guide to the eye and open symbols depict transition temperatures from wTF measurements.
b) ``Quantum melting'' of magnetic order: saturation field $B_{\rm m}$ at $0.24\,{\rm K}$ for different disorder concentrations and pressures at 9\,kbar and 22\,kbar (for 4\% the data are for 7~kbar).}
}
\end{figure}

In Fig.~\ref{fig:EstimationSaturationField} we show the fitted $B_{\rm m}$ against temperature\footnote{Interestingly, the ordering temperature of 4\% PHCX at 18\,kbar is lower than the one of 1\% PHCX at 9\,kbar. At the same time $B_{\rm m}$ at low temperatures is higher for the 4\% PHCX at 18\,kbar than for 1\% PHCX at 9\,kbar. However, this behavior is not related to disorder and the same applies for pure PHCC around the Lifshitz point. The effect remains to be fully understood, for instance by computational studies of PHCX under pressure.} and disorder concentration. For either pressure, $B_{\rm m}$ drops with increasing disorder concentration. This drop is far too great to be explained by the small increase of lattice parameters as a function of increased level of disorder. Hence $B_{\rm m}$ is primarily a measure of the ordered moment size in the sample. The measurement shows that the ordered moments are reduced with increasing disorder concentration until ordering is suppressed completely at 15\% Br-concentration for all investigated pressures.

\subsection{Weak-transverse fields}
\begin{figure}
\unitlength1cm
\includegraphics[width=\columnwidth]{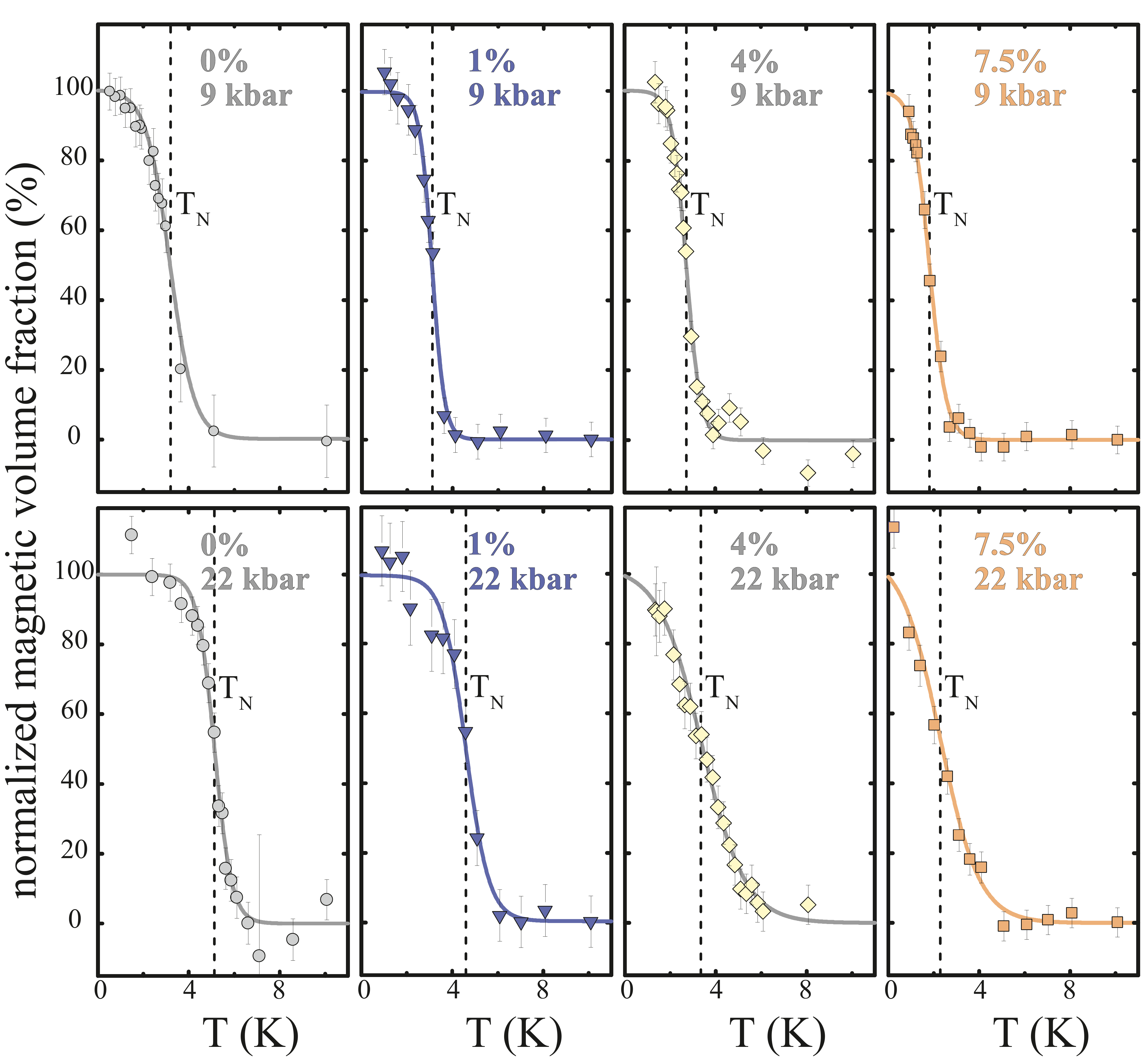}
\caption{\label{fig:TempDependence}
Temperature dependence of the normalized magnetic volume fraction of PHCX in weak transverse magnetic fields of $30\,{\rm mT}$, assuming a fully ordered sample at low temperatures. Data are shown for pure PHCC and Br-concentrations of $x= 1\%$, $4\%$ and $7.5\%$, each at intermediate and at high pressures of $9\,{\rm kbar}$ (top) and $22\,{\rm kbar}$ (bottom), respectively. Lines are sigmoidal fits to the data to determine the transition temperatures $T_{\rm N}$ (dashed lines) at the steepest slope.}
\end{figure}

\begin{figure}
\unitlength1cm
\includegraphics[width=\columnwidth]{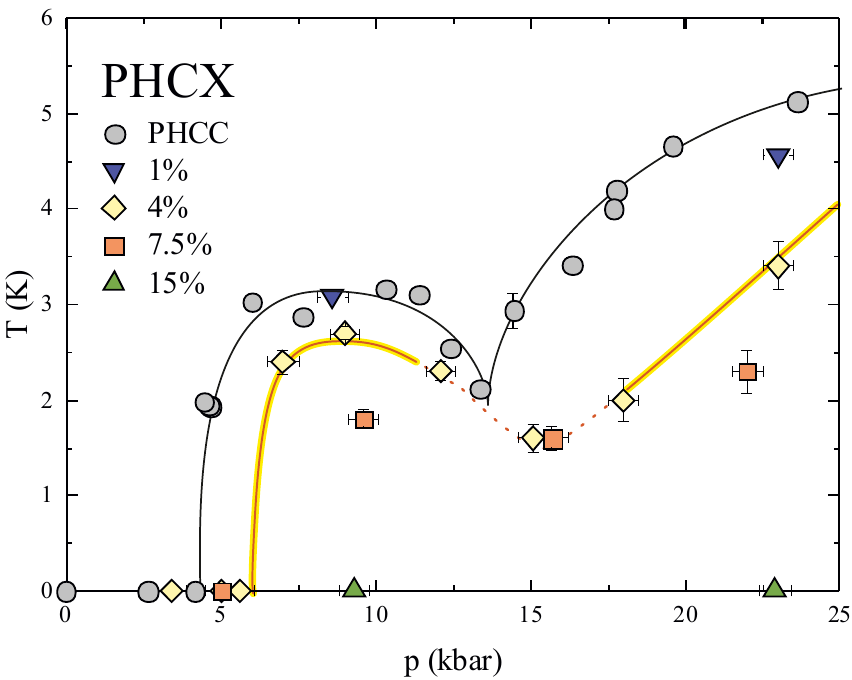}
\caption{\label{fig:PhaseDiagrams}
$p$--$T$ phase diagram of PHCX for Br-concentrations of $x=1\%$, $4\%$, $7.5\%$ and $15\%$. Depicted in grey are the data of the disorder-free compound PHCC~\cite{Thede2014PRL}. Symbols are transition temperatures as determined by wTF measurements. Symbols at $T\,=\,0$~K indicate the absence of magnetic order down to 0.24~K. Lines are a guide to the eye.}
\end{figure}

\musr\ in weak transversal fields (wTF) was used to determine transition temperatures and map out the phase diagram of PHCX for a number of disorder concentrations with a procedure analogous to that used for disorder-free PHCC~\cite{Thede2014PRL}. The temperature dependence of the normalized magnetic volume fraction is plotted in Fig.~\ref{fig:TempDependence} for 1\%-, 4\%- and 7.5\%-disordered PHCX. For comparison the corresponding plots for pure PHCC have been added. The transition from ordered to paramagnetic state is marked by a sudden drop in the magnetic volume fraction. The transition temperatures were determined by sigmoidal fits to the temperature dependence of the magnetic volume fraction. Based on ZF data and a comparison with PHCC it can be reasonably assumed that at low temperatures all samples with the exception of 15\%-disordered PHCX are magnetically ordered throughout the bulk of the sample, regardless of disorder concentration and pressure. However, an accurate determination of the magnetic volume fraction is not possible due to the uncertainty in the fraction of muons stopped in the pressure cell, which depends on the beamline parameters and the properties of the sample. The wTF transition widths increase slightly with increasing pressure. From Fig.~\ref{fig:TempDependence}, it becomes immediately evident that transition temperatures at both pressures decrease with increasing disorder concentration. Furthermore, for high levels of disorder, the ordering temperatures are less sensitive to the applied pressure than at lower concentrations. These data are summarized in the $p$-$T$ phase diagram of PHCX for all measured disorder concentrations and for pure PHCC which is shown in Fig.~\ref{fig:PhaseDiagrams}.

\section{Discussion}

Our data show that at least up to nominal Br-concentration of 7.5\%, static magnetic order can be induced in PHCX by application of hydrostatic pressure of up to 22~kbar. Indeed, the $p$--$T$ phase diagram obtained for PHCX with $x\leq7.5\%$ qualitatively resembles that of pure PHCC. However, compared to pure PHCC, even small levels of disorder lead to a heavy damping of the \musr\ oscillations observed in ZF in the ordered phases, indicating a more inhomogeneous magnetic phase (Fig.~\ref{fig:PHCX_diffBrConcentr}). Furthermore we find that the ordering temperatures and ordered moment sizes are suppressed by increasing levels of disorder. The critical pressure for ordering is in the range $6~{\rm kbar}<p_{\rm c}<7$~kbar for 4\%-disordered PHCX compared with $p_{\rm c}\sim 4.3$~kbar for pure PHCC, corresponding to an approximately 50\% increase in critical pressure. It is interesting to remark that inelastic neutron scattering found that the gap in Br-disordered PHCX increases to 1.36~meV and 1.50~meV for 3.5\% and 7.5\%-disordered samples compared to 1.0~meV for pure PHCC~\cite{Dan2012_gap}. It is hence plausible that the critical pressure is increased at 4\% Br-concentration compared to pure PHCC; the magnitude of the shift is also consistent. Finally, at 15\%, pressure-induced order is completely destroyed. This may be due to disorder-induced frustration that leads to a large number of quasi-degenerate ground state configurations and thus the suppression of magnetic order. Similar effects were observed at comparable disorder concentrations for PHCX around the $z=2$ QCP in applied magnetic fields~\cite{Dan2012}. This process can be thought of as ``quantum melting'' of magnetic order analogous to the thermal melting observed as a function of temperature. 

Disorder has a negligible effect on a classical antiferromagnet. Therefore, Br-substitution would only be expected to significantly affect the magnetism in PHCX due to quantum effects. Bond disorder may cause a spatially random variation of the strength of local quantum spin fluctuations, which results in the inhomogeneous spatial distribution of ordered moments. Such an effect was observed by \musr~experiments in bond-disordered spin chains~\cite{Thede2014}.

Recent numerical work has predicted the existence of a strongly inhomogeneous quantum Griffiths regime in bond-disordered dimer magnets near quantum criticality~\cite{Vojta2013}. This numerical study considered a bilayer Heisenberg dimer system in which dimers are located on a two-dimensional square lattice, and the effect of randomly-modifying the interlayer/intradimer coupling $J$ and interdimer/intralayer coupling $K$ was investigated. The interdimer coupling was used to tune the system around the $z=1$ QCP allowing detailed predictions for the magnetic excitations in this regime. Finite islands of nonzero staggered magnetization with an exponentially large distribution of fields appear in a Griffiths phase~\cite{Griffiths} below the critical interdimer coupling $K_{\rm c}$. It is tempting to draw a direct analogy between pressure and the tuning parameter $K$, and Br-concentration $x$ and the degree of disorder. However, we must exercise some caution and we note that we did not observe an onset of order in PHCX below the critical pressure of pure PHCC, but instead the critical pressure is \emph{increased} by Br-substitution. Even for pure PHCC there are at least six different relevant exchange interactions~\cite{Stone2001}. The system is therefore intrinsically more complicated than the model dimer system considered and it is known that bond-disorder affects several bonds with varying substitution rates~\cite{Dan2013_xray}. In any real system substituting the ligand not only introduces bond disorder but also chemical pressure which may alter some of the other parameters in the Hamiltonian as well. Therefore there may be two competing effects in the system: (i) a decrease in the gap by introducing disorder as studied theoretically and (ii) an increase in the gap by chemical pressure, which dominates here. 

We note that the behavior we observed in PHCX is qualitatively very similar to that found using \musr\ in disordered itinerant Ni$_{1-x}$V$_x$ alloys~\cite{Schroeder2014,Wang2015}. In Ni$_{1-x}$V$_x$ a Griffiths phase has been revealed by bulk measurements\citep{Ubaid2010} above a critical disorder concentration $x_{c}=11.4\%$. For $x=0$ sharply defined oscillations are observed in the muon asymmetry indicating homogeneous long-range magnetic order. At intermediate concentrations $0<x<x_{\rm c}$ increasingly inhomogeneous behavior evidenced by strongly damped oscillations is observed and ordered moment sizes and ordering temperatures are suppressed. In this phase Ni$_{1-x}$V$_x$ is an inhomogeneous ferromagnet. Then above a QCP at $x_{\rm c}=11.4\%$, no oscillations are observed in ZF and wTF measurements show no change as a function of temperature as the system enters a quantum Griffiths phase. Qualitatively this is precisely what we observe in PHCX: for $x=0$ sharp oscillations indicate homogeneous magnetic order above the critical pressure. For small disorder concentrations $0<x<7.5\%$ the oscillations observed above a critical pressure become increasingly damped, while moment sizes and ordering temperatures decrease with Br-concentration. Finally, at $x=15\%$, magnetic order can no longer be achieved at up to 22~kbar applied pressure. Because of this analogy we believe that the formation of an inhomogeneous Griffiths phase is a possible cause of the observed behavior in PHCX. PHCX would then enter a potential quantum Griffiths phase in the range $7.5\% < x_{c} < 15\%$.
We note that at low temperatures there is a freezing into a cluster glass phase in Ni$_{1-x}$V$_x$ for which we have observed no evidence in PHCX.

Inelastic-neutron scattering experiments are required to directly search for the predicted in-gap states and disorder-induced magnon broadening in the predicted Griffiths phase. At ambient pressure an energy-dependent broadening of the magnons was observed in 3.5\% and 7.5\%-disordered PHCX~\cite{Dan2012_gap}, which was shown to be caused by single-magnon scattering by impurities and is therefore not related to the Griffiths phase physics discussed above. This broadening has also been observed in electron-spin resonance experiments on Br-disordered PHCX~\cite{Glazkov2014}.

\section{Conclusion}

In summary, we have studied the behavior of the bond-disordered quantum magnet PHCX under hydrostatic pressure. We find that for Br-concentrations of $x\leq7.5\%$ PHCX undergoes a pressure-induced quantum phase transition into a magnetically ordered state with a $p$--$T$ phase diagram that broadly resembles that of pure PHCC. However, even for small Br-concentrations $x>1\%$ the pressure-induced ordered phases display highly inhomogeneous magnetism. Ordering temperatures and ordered moments are suppressed by increasing disorder levels and the critical pressure increases. At 15\% Br-concentration magnetic order is not detectable. Qualitatively the behavior of PHCX under hydrostatic pressure at high disorder concentrations is consistent with recent predictions for a Griffiths phase region in bond-disordered dimer magnets close to quantum criticality.


\acknowledgments
We thank Almut Schroeder for helpful discussions.
This work was supported by the Swiss National Science Foundation. 
J.S.M.\  is grateful for support by the ETH Zurich Postdoctoral Fellowship Program which has received funding from the European Union's Seventh Framework Programme for research, technological development and demonstration under grant agreement 246543. 
This project has received funding from the European Union's Seventh Framework Programme for research, technological development and demonstration under the NMI3-II Grant number 283883. The UK-based authors are supported by EPSRC (UK). The data shown in this paper is available at \url{http://link.aps.org/supplemental/10.1103/PhysRevB.94.144418}.
This work was partly supported by the Estonian Ministry of Education and Research under grant No. IUT23-03, the Estonian Research Council grant No. PUT451 and the European Regional Development Fund project TK134.
The experiments were performed on the GPD beamline at the Swiss muon source at Paul Scherrer Institut, Switzerland.

\bibliography{../../Literature/BiBTeX}

\end{document}